\documentclass[sigconf]{acmart}

\AtBeginDocument{%
  }

\setcopyright{acmlicensed}
\acmDOI{XXXXXXX.XXXXXXX}
\usepackage[table]{xcolor}
\usepackage{hyperref}
\usepackage{url}
\usepackage{multirow}
\usepackage{colortbl}
\usepackage{soul}
\usepackage{subcaption}
\usepackage{makecell}
\usepackage{pifont}
\newcommand{\cmark}{\ding{51}}
\newcommand{\xmark}{\ding{55}}

\definecolor{lightgreen}{RGB}{173, 216, 230}
\definecolor{lightred}{RGB}{255, 218, 185}
\definecolor{MyLightRed}{RGB}{255,150,150}
\definecolor{MyDarkRed}{RGB}{139,0,0}
\begin{document}

\def\method{\textsc{ReAlign}}

\title{\method{}: Optimizing the Visual Document Retriever with Reasoning-Guided Fine-Grained Alignment}

\author{Hao Yang}
\authornote{ \ \ indicates equal contribution.}
\affiliation{%
  \institution{Northeastern University}
  \city{Shenyang}
  \country{China}}
\email{yanghao123@mails.neu.edu.cn}

\author{Yifan Ji}
\authornotemark[1]
\affiliation{%
  \institution{Northeastern University}
  \city{Shenyang}
  \country{China}}
\email{jiyf1@mails.neu.edu.cn}

\author{Zhipeng Xu}
\authornotemark[1]
\affiliation{%
  \institution{Northeastern University}
  \city{Shenyang}
  \country{China}}
\email{xuzp@mails.neu.edu.cn}

\author{Zhenghao Liu}
\authornote{ \ \ indicates corresponding author.}
\affiliation{%
  \institution{Northeastern University}
  \city{Shenyang}
  \country{China}}
\email{liuzhenghao@mail.neu.edu.cn}

\author{Yukun Yan}
\affiliation{%
 \institution{Tsinghua University}
 \city{Beijing}
 \country{China}}
\email{yanyk.thu@gmail.com}

\author{Zulong Chen}
\affiliation{%
  \institution{Alibaba Group}
  \city{Hangzhou}
  \country{China}}
\email{zulong.czl@alibaba-inc.com}

\author{Shuo Wang}
\affiliation{%
  \institution{Tsinghua University}
  \city{Beijing}
  \country{China}}
\email{wangshuo.thu@gmail.com}

\author{Yu Gu}
\affiliation{%
  \institution{Northeastern University}
  \city{Shenyang}
  \country{China}}
\email{guyu@mail.neu.edu.cn}

\author{Ge Yu}
\affiliation{%
  \institution{Northeastern University}
  \city{Shenyang}
  \country{China}}
\email{yuge@mail.neu.edu.cn}

\renewcommand{\shortauthors}{Hao Yang et al.}

\begin{abstract}
Visual document retrieval aims to retrieve a set of document pages relevant to a query from visually rich collections. Existing methods often employ Vision-Language Models (VLMs) to encode queries and visual pages into a shared embedding space, which is then optimized via contrastive training. However, during visual document representation, localized evidence is usually scattered across complex document layouts, making it difficult for retrieval models to capture crucial cues for effective embedding learning. In this paper, we propose Reasoning-Guided Alignment (\method{}), a method that enhances visual document retrieval by leveraging the reasoning capability of VLMs to provide fine-grained visual document descriptions as supervision signals for training. Specifically, \method{} employs a superior VLM to identify query-related regions on a page and then generates a query-aware description grounding the cropped visual regions. The retriever is then trained using these region-focused descriptions to align the semantics between queries and visual documents by encouraging the document ranking distribution induced by the region-focused descriptions to match that induced by the original query. Experiments on diverse visually rich document retrieval benchmarks demonstrate that \method{} consistently improves visual document retrieval performance on both in-domain and out-of-domain datasets, achieving up to 2\% relative improvements. 
Moreover, the advantages of \method{} generalize across different VLM backbones by guiding models to better focus their attention on critical visual cues for document representation. All code and datasets are available at \url{https://github.com/NEUIR/ReAlign}.
\end{abstract}




\maketitle

\section{Introduction}
Visual document retrieval aims to identify document pages relevant to a given query from large collections of visually rich documents~\cite{takeda2011real,zhalehpour2019visual,giotis2017survey,faysse2025colpali}. 
It usually serves as a fundamental component for various downstream document understanding tasks, including document question answering~\cite{tanaka2023slidevqa}, fact verification~\cite{schuster2019towards,bekoulis2021review}, and information extraction~\cite{aumann2006visual,gao2012view}. 
Despite its importance, visual document retrieval remains challenging due to the inherent complexity of document images~\cite{marinai2011digital,guo2025towards}. 
Unlike natural images, document pages present highly heterogeneous layouts that are tightly coupled with textual content, with content often sparsely scattered across multiple regions~\cite{xu2020layoutlm,appalaraju2021docformer,yu2024texthawk,li2025regionrag}.
Although visual documents contain richer semantics, the query-document relevance is usually determined by a small number of localized regions, such as specific fields, headings, or key-value pairs, while the majority of page content is irrelevant and may even introduce misleading signals~\cite{wen2023visual,cao2023attention,li2025regionrag}. 
Thus, visual document retrieval requires models to understand complex layout structures and effectively capture some necessary evidence from the entire page~\cite{faysse2025colpali,mace2025vidore,yuan2023vile}.

\begin{figure}[t]
\centering
\includegraphics[width=\linewidth]{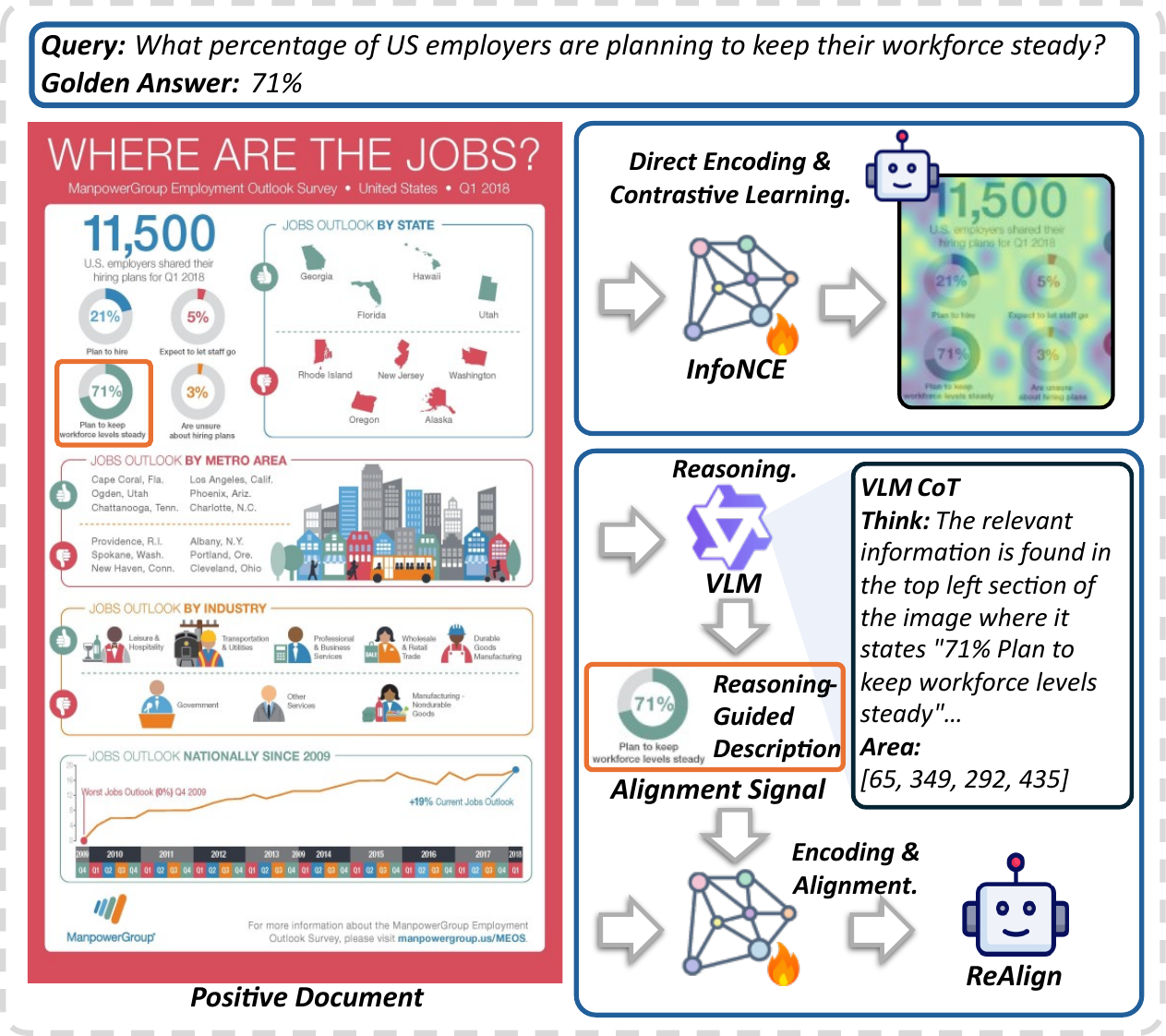}
\caption{Illustration of Our Reasoning-Guided Alignment (\method{}) Method for Visual Document Retrieval. The orange box represents the ground-truth region.\label{fig:intro}}
\end{figure}
To address this problem, recent research relies on the strong capabilities of Vision-Language Models (VLMs), directly encoding document pages into embeddings and adopting contrastive training objectives to align queries and visual documents through relevance modeling~\cite{ma2024unifying,tanaka2025vdocrag,yu2025visrag}. 
While effective, these approaches often struggle to accurately capture fine-grained visual cues during representation learning when relying solely on contrastive objectives such as InfoNCE~\cite{oord2018representation}. 
As illustrated in Figure~\ref{fig:intro}, the attention distribution produced by InfoNCE-based training tends to be diffusely spread around the boundaries of ground-truth regions, rather than concentrating on the truly critical visual evidence.
To encourage VLMs to better focus on salient visual evidence, recent works leverage the reasoning capabilities of VLMs by prompting them to interact with auxiliary image tools, such as zoom-in and zoom-out operations, enabling more precise localization of query-relevant regions in visual documents~\cite{wang2025pixel,shen2025zoomeye,wang2025vidorag,wang2025vrag}. 
By exploiting these reasoning processes, the model can accurately localize the target value ``71\%'' using zoom-in results with explicit bounding box coordinates, thereby providing finer-grained supervisory signals. 
Such signals are beneficial for guiding VLMs to achieve better semantic alignment between queries and documents during training.

In this paper, we propose the Reasoning-Guided Alignment (\method{}) method, a framework that leverages the visual reasoning capabilities of VLMs to uncover some query-relevant evidence within document pages. This process provides fine-grained supervision signals to guide the training of visual document retrievers.
Specifically, \method{} employs a high-capacity VLM to perform a reasoning process that localizes query-related regions in the given visual documents. Based on the grounded bounding box coordinates of the identified regions, the VLM is then prompted to generate visual document descriptions. Besides query-document relevance, these region-aware descriptions serve as additional supervision signals to guide the visual document representation learning of VLMs. During training, the query-document relevance reflects the ground-truth user intent, and the model is optimized to minimize the discrepancy between the description-induced ranking probability and the ranking distribution derived from query-document relevance. In this way, the region-focused description functions as a regularization objective, enhancing fine-grained semantic alignment between the query and the corresponding visual document.

Our experiments on multiple visual document retrieval benchmarks demonstrate that \method{} yields significant performance gains over baseline models, validating its overall effectiveness.
Moreover, \method{} consistently outperforms baseline approaches across different VLM backbones, highlighting its strong generalization capability.
Further analysis shows that \method{} effectively guides the retriever to learn a more discriminative embedding space by aligning queries with their corresponding visual documents, while simultaneously maintaining embedding space uniformity to better distinguish different visual documents. 
During training with \method{}, the VLM is optimized to allocate greater attention to query-relevant regions identified through the reasoning process of superior VLMs. After training, the visual document retriever learns to more effectively encode semantic information from query-relevant regions and to capture critical evidence, such as numerical cues, within visual document representations.
This attention mechanism enables the visual document retriever to achieve superior retrieval performance, particularly in challenging scenarios where relevance depends on localized visual evidence within complex document layouts.
\section{Related Work}
Visual document retrieval is a fundamental problem in document understanding, which aims to identify document pages relevant to a given query from large collections of visually rich documents~\cite{doermann1998indexing, marinai2011digital, alaei2016brief}.
Early studies predominantly rely on Optical Character Recognition (OCR) to transform visual document pages into plain text~\cite{alaei2016document,ahmed2017survey,zhang2025ocr,guo2025towards}, thereby reducing visual document retrieval to a conventional text retrieval setting, where standard text-based retrieval models are directly employed to rank documents~\cite{karpukhin2020dense,zagoris2010document,ji2025learning}.
Although effective in practice, such approaches are highly sensitive to the quality of OCR outputs, which often introduces unnecessary cascading errors into downstream retrieval models~\cite{bazzo2020assessing,zhang2025ocr,shim2025revise,mei2018statistical,song2026defining}.
Furthermore, text extracted by OCR systems fails to faithfully preserve the original layout and spatial organization of document pages, frequently weakening or discarding layout cues that are crucial for accurate document retrieval~\cite{keyvanpour2013document,li2021structext,appalaraju2024docformerv2,xu2020layoutlm}.
As a result, OCR-based methods often struggle to robustly model the rich visual and layout information inherent in document pages, limiting their effectiveness in scenarios where retrieval relevance critically depends on layout-aware and spatially grounded evidence~\cite{powalski2021going,wang2022lilt,peng2022ernie}.

More recent efforts~\cite{yu2025visrag,faysse2025colpali,tanaka2025vdocrag,ma2024unifying} have explored adapting Vision-Language Models (VLMs) to directly encode visual documents into a shared embedding space for retrieval, and to estimate the relevance between queries and visual document pages by computing their similarity scores~\cite{sun2025unveil,ke2025large,kim2022ocr,liu2024textmonkey}. Benefiting from the strong emergent capabilities of VLMs~\cite{wei2022emergent,zhao2023survey}, some works~\cite{li2024llama2vec,jiang2025vlm2vec} directly prompt VLMs to produce unified representations for both queries and documents, enabling end-to-end retrieval modeling.
Furthermore, to enhance the representation capability of VLMs, some research follows the contrastive training paradigm in dense retrieval~\cite{karpukhin2020dense,izacard2021unsupervised}, training retrievers by aligning document and query representations using global page-level supervision~\cite{ma2024unifying,yu2025visrag}.
Despite these methods showing effectiveness in retrieval by avoiding unnecessary OCR errors through end-to-end document page retrieval, such supervision remains coarse-grained, providing limited guidance on which specific visual or textual elements within a document actually support the relevance judgment~\cite{teiletche2025modernvbert,cui2025attention,li2024visual}.

To mitigate this issue, recent studies have focused on enhancing the fine-grained perceptual capacity of visual document retrievers, thereby enabling more localized evidence modeling~\cite{tong2025hkrag,li2025regionrag}.
VDocRetriever~\cite{tanaka2025vdocrag} trains VLMs to learn encoded representations of visual document pages by reproducing OCR results and aligning image representations with the corresponding textual representations derived from OCR.
ColPali~\cite{faysse2025colpali} further partitions each document page into multiple visual regions and performs matching between query tokens and these regions, aggregating region-level similarity scores to estimate relevance based on localized alignments rather than a single global representation.
While these approaches improve fine-grained perceptual modeling, they still rely on indirect supervision and do not explicitly specify which localized evidence grounds the query relevance.
As a result, the learned representations lack explicit evidence grounding, which hampers robust identification of query-relevant regions in complex document layouts~\cite{liu2025look}.

To enhance the visual perception capabilities of VLMs, recent works have leveraged their inherent reasoning ability to enable reasoning-guided visual focusing behaviors, where models dynamically attend to query-relevant regions through implicit visual exploration during the reasoning process~\cite{wang2025pixel,shen2025zoomeye,wang2025vidorag,wang2025vrag}.
DyFo~\cite{li2025dyfo} introduces dynamic focusing by continuously updating attended regions during reasoning, allowing attention to adaptively shift across different regions of the input.
In contrast, Chain-of-Focus~\cite{zhang2025chain} explicitly formulates reasoning-guided focusing as a coarse-to-fine process, in which attention is progressively narrowed and refined in alignment with intermediate reasoning states.
PixelReasoner~\cite{wang2025pixel} further advances this line of work by explicitly modeling pixel-level visual operations, such as zooming and region selection, and integrating them into multi-step reasoning processes, thereby enabling the model to make explicit decisions about where to attend.
Collectively, these recent advances suggest that leveraging reasoning signals to guide visual attention provides a more principled mechanism for localizing query-relevant evidence in complex documents~\cite{shih2016look,kang2025your,lu2025multimodal,li2025towards}.
However, existing retrievers primarily rely on global alignment signals between the query and entire document pages~\cite{yu2025visrag,ma2024unifying,bakkali2025globaldoc}. In contrast, the reasoning-guided focusing capabilities remain largely unexplored and have not yet been incorporated as fine-grained supervision signals for optimizing visual document retrieval.
\section{Methodology}
In this section, we first introduce the preliminaries of visual document retrieval (Sec.~\ref{method:preliminary}), and then present the reasoning-guided alignment mechanism adopted in \method{} (Sec.~\ref{method:Reasoning-Guided-Alignment}).

\subsection{Preliminaries of Visual Document Retrieval}
\label{method:preliminary}
Given a query $q$ and a visually rich document collection $\mathcal{D}=\{ d_1, \ldots, d_n \}$, where each document $d$ corresponds to an image of a single document page, the goal of visual document retrieval is to retrieve a set of documents from the collection that are relevant to the query.

Specifically, VLM-based visual document retrievers leverage a Vision-Language Model (VLM) $\mathcal{M}$ to encode the query $q$ and a document $d$ into dense embeddings $E_q$ and $E_d$, respectively:
\begin{equation}
E_q = \mathcal{M}(q), \quad E_d = \mathcal{M}(d).
\end{equation}
The relevance score $f(q, d)$ between the query embedding $E_q$ and the document embedding $E_d$ is then defined as:
\begin{equation}
\label{eq:score}
f(q, d) = sim(E_q, E_d),
\end{equation}
where \textit{sim} denotes a similarity function. In \method{}, cosine similarity is employed to measure the semantic similarity between the query and the document embeddings. The query encoder and document encoder are trained in a contrastive manner by maximizing the ranking probability $P(d^{+}\mid q, \{d^{+}\}\cup\mathcal{D}^{-})$ of the query-related visual document $d^{+}$:
\begin{equation}
\label{eq:retrieval_prob}
P(d^{+}\mid q,\{d^{+}\}\cup\mathcal{D}^{-})
=
\frac{e^{f(q,d^{+})}}
{e^{f(q,d^{+})}+\sum_{d^{-}\in\mathcal{D}^{-}} e^{f(q,d^{-})}},
\end{equation}
where $d^{-}$ denotes a document sampled from the irrelevant document set $\mathcal{D}^{-}$~\cite{karpukhin2020dense,xiong2020approximate}, such as in-batch negatives.

\begin{figure*}[t]
\begin{center}
\includegraphics[width=1\linewidth]{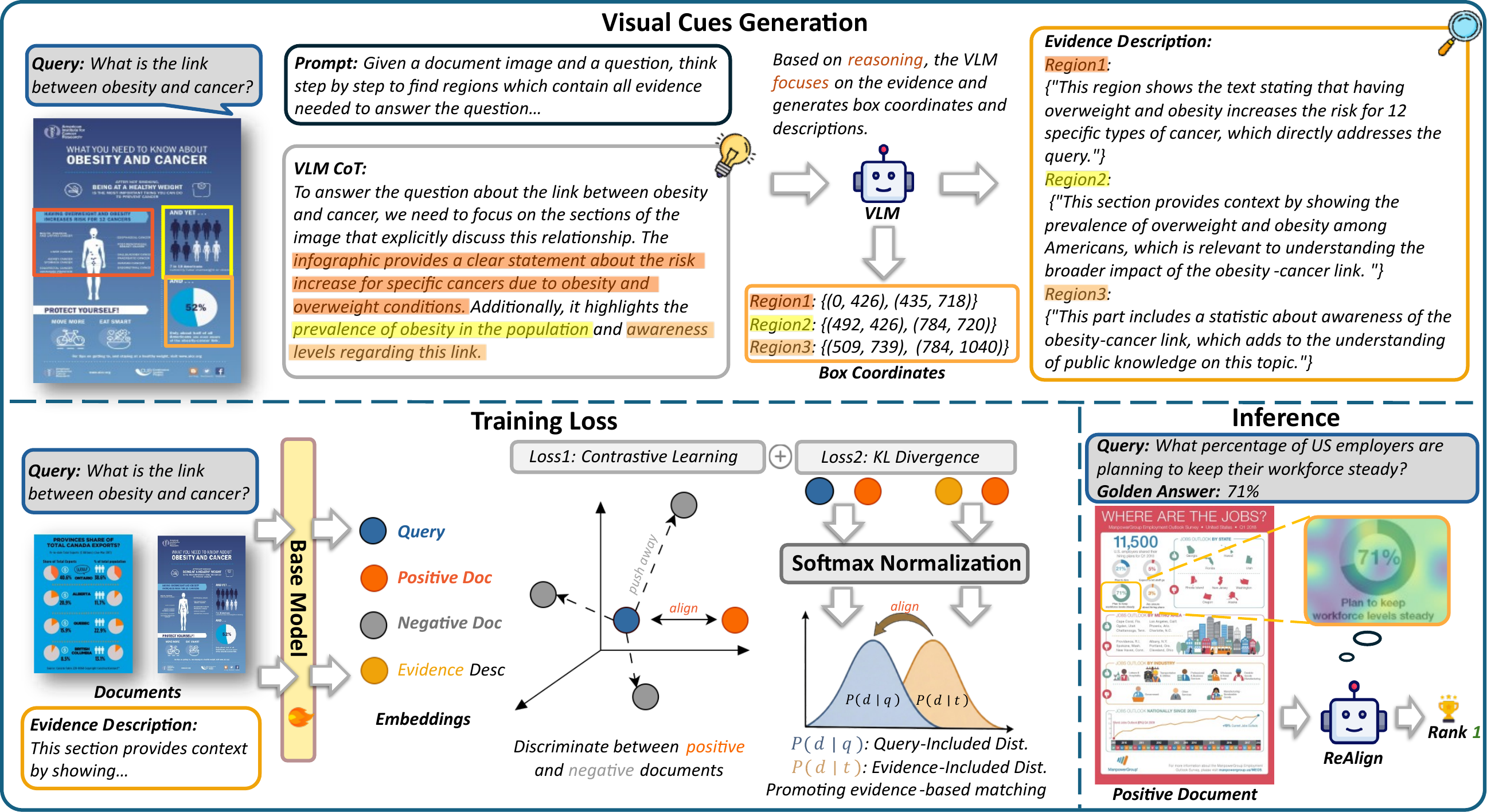}
\end{center}
\caption{The Architecture of Reasoning-Guided Visual Document Retrieval (\method).}
\label{fig:main_figure}
\end{figure*}
\subsection{\method{}: Reasoning-Guided Fine-Grained Visual-Language Alignment}
\label{method:Reasoning-Guided-Alignment}
As shown in Figure~\ref{fig:main_figure}, we introduce \method{} to provide additional fine-grained visual-language alignment signals for training visual document retrievers with the query-document pairs.

Given a query-document pair $(q,d)$, existing works~\cite{faysse2025colpali,kolouju2025good4cir,nguyen2025serval} typically ask VLMs $\mathcal{M}$ to ground the visual document and generate a corresponding textual description $t$ that verbalizes the document image:
\begin{equation}
t =\mathcal{M}(d),
\label{eq:cue_gen_typically}
\end{equation}
where $(t,d)$ is treated as supervision for continuously pretraining VLMs, enabling them to better represent both queries and images by bridging the modality gap through generative objectives~\cite{liu2023univldr}.
Although effective, such approaches primarily focus on global visual semantics and fail to encourage VLMs to capture subtle and fine-grained cues in visual documents~\cite{wang2025pixel}, particularly the query-relevant regions within the document images. As a result, the learned visual document representations remain coarse-grained, which limits their effectiveness in fine-grained retrieval scenarios.
To address this limitation, \method{} synthesizes fine-grained supervision signals that explicitly guide VLMs to capture subtle semantics from query-document pairs $(q,d)$ during SFT. In the remainder of this subsection, we first describe the supervision synthesis process, and then present how these signals are leveraged to optimize the visual document retriever.

\textbf{Region-Guided Supervision Syntheses.}
To facilitate VLMs in better understanding the semantics of the visual document $d$ during training, we leverage the reasoning capability of the VLM $\mathcal{M}$ to synthesize auxiliary supervision signals. These signals are designed to help VLMs more effectively align the query $q$ with its corresponding document $d$.

Specifically, we first prompt the VLM $\mathcal{M}$ to identify $K$ query-related regions from the visual document $d$, which encourages the model to attend to these regions during training:
\begin{equation}
{(b_i,t_i)}_{i=1}^{K}=\mathcal{M}(q,d),
\label{eq:cue_gen}
\end{equation}
where $b_i$ and $t_i$ denote the localized region in the visual document $d$ and its corresponding evidence description, respectively. Each region $b_i$ is represented by the coordinates of a bounding box $[x_1, y_1, x_2, y_2]$, where $(x_1, y_1)$ and $(x_2, y_2)$ correspond to the top-left and bottom-right corners of the bounding box, respectively. The bounding box coordinates serve as prompts that guide the VLM to focus on the specified regions of the visual document $d$ when generating the region-focused description $t_i$.

To ensure the diversity of the synthesized supervision signals and avoid information redundancy, we randomly sample one description $t$ from the candidate set for each query:
\begin{equation}
t \sim \mathcal{U}(\mathcal{T}), \quad \mathcal{T} = \{t_k\}_{k=1}^{K}.
\label{eq:sample}
\end{equation}
Finally, we construct the training dataset by pairing the query $q$, the visual document $d$, and the sampled region-focused description $t$, forming a triplet $(q, d, t)$ for model optimization.

\textbf{Reasoning-Guided Vision-Language Alignment.}
After collecting all query-document-description triplets $(q, d, t)$, we propose the ranking distribution alignment method, which leverages the region-focused description $t$ to help the VLM better learn both query and visual document representations.

Specifically, for each training instance consisting of a query $q$, a relevant document $d^{+}$, and a set of irrelevant documents $\mathcal{D}^{-}$, we construct the candidate set:
\begin{equation}
\tilde{\mathcal{D}}=\{d^{+}\}\cup\mathcal{D}^{-}.
\end{equation}
We then compute the query-induced retrieval distribution $P(d\mid q,\tilde{\mathcal{D}})$ and the evidence-induced retrieval distribution $P(d\mid t,\tilde{\mathcal{D}})$ using Eq.~\ref{eq:retrieval_prob}.
To align these two distributions, we employ the KL divergence as a regularization objective, encouraging the evidence-induced distribution to approximate the query-induced distribution:
\begin{equation}
\mathcal{L}_\text{KL}=\sum_{q}\ \sum_{d\in\tilde{\mathcal{D}}} P(d\mid q,\tilde{\mathcal{D}}) \cdot \log\frac{P(d\mid q,\tilde{\mathcal{D}})}{P(d\mid t,\tilde{\mathcal{D}})}.
\end{equation}
This alignment objective enforces distributional consistency between the query $q$ and its evidence description $t$. The query-induced distribution $P(d\mid q,\tilde{\mathcal{D}})$ acts as a teacher signal, as it directly captures the retrieval intent under explicit supervision, thereby guiding the VLMs to attend to fine-grained visual evidence for the description-document matching, rather than relying on coarse-grained global similarity.

Finally, we optimize our \method{} using the training objective:
\begin{equation}\label{eq:training_loss}
\mathcal{L}=\mathcal{L}_\text{Contrast}+\lambda\,\mathcal{L}_\text{KL},
\end{equation}
where $\mathcal{L}_\text{Contrast}$ denotes the standard contrastive learning loss over the query-document pair $(q, d)$ that maximizes the retrieval probability defined in Eq.~\ref{eq:retrieval_prob}, and $\lambda$ is a hyper-parameter that balances the contrastive objective and the proposed distribution alignment regularization.

\section{Experimental Methodology}

In this section, we introduce the datasets, evaluation metrics, baselines, and implementation details of our experiments.

\begin{table}[t]
        \centering
        \caption{Training Dataset Statistics.}
        \label{tab:train_stats}

        \resizebox{\columnwidth}{!}{%
        \begin{tabular}{llrrr}
            \hline
            Dataset & Field & \#Images & \#Query & \#Desc \\ 
            \hline
            DocVQA~\cite{mathew2021docvqa}        & Industry    & 12,767 & 6,382 & 6382 \\
            InfoVQA~\cite{mathew2022infographicvqa}& Infographic & 5,485  & 9,592 & 9,587 \\
            VisualMRC~\cite{tanaka2021visualmrc}  & Webpage     & 10,229 & 6,126 & 6,120 \\
            OpenWikiTable~\cite{kweon2023open}& Table       & 1,257  & 4,261 & 4248 \\
            DUDE~\cite{van2023document}            & Open        & 27,955 & 2,135 & 2043 \\
            MHDocVQA~\cite{tanaka2025vdocrag}                    & Open        & 28,550 & 9,470 & 80 \\
            \hline
        \end{tabular}
        }
\end{table}

\textbf{Datasets.} 
We follow the experimental setting of \citet{tanaka2025vdocrag} to conduct our experiment.
The training set comprises approximately 38,000 query-document pairs sampled from DocVQA~\cite{mathew2021docvqa}, InfoVQA~\cite{mathew2022infographicvqa}, VisualMRC~\cite{tanaka2021visualmrc}, OpenWikiTable~\cite{kweon2023open}, DUDE~\cite{van2023document}, and MHDocVQA~\cite{tanaka2025vdocrag}, with dataset statistics shown in Table~\ref{tab:train_stats}. Note that MPMQA~\cite{zhang2023mpmqa} is excluded as it is not available in their official repository. For evaluation, we test the proposed retriever on six visual document retrieval benchmarks, including in-domain evaluations on DocVQA and InfoVQA, as well as zero-shot evaluations on ChartQA~\cite{masry2022chartqa}, SlideVQA~\cite{tanaka2023slidevqa}, PlotQA~\cite{methani2020plotqa}, and ArXivQA~\cite{li2024multimodal}. Detailed statistics are reported in Table~\ref{tab:eval_dataset_stats}.

\begin{table}[t]
\centering
\caption{Test Dataset Statistics.}
\label{tab:eval_dataset_stats}
\begin{tabular}{l l r r c}
\hline
Dataset     & Field         & \#Images  & \#Query  &Zero-Shot \\
\hline
DocVQA~\cite{mathew2021docvqa}      & Industry      & 741       & 585      & \xmark  \\
InfoVQA~\cite{mathew2022infographicvqa}     & Infographic   & 5,485     & 1,048    & \xmark   \\
ChartQA~\cite{masry2022chartqa}     & Open          & 20,882    & 150      & \cmark  \\
SlideVQA~\cite{tanaka2023slidevqa}    & Open          & 52,380    & 760      & \cmark  \\
PlotQA~\cite{methani2020plotqa}      & Scientific    & 9,593     & 863      & \cmark  \\
ArXivQA~\cite{li2024multimodal}     & Academic      & 8,066     & 816      & \cmark  \\
\hline
\end{tabular}
\end{table}

\textbf{Evaluation Metrics.}
To assess the effectiveness of \method{}, we adopt NDCG@5 and NDCG@10 as the evaluation metrics, following prior work~\cite{yu2025visrag,tanaka2025vdocrag}. NDCG scores are computed using the official implementation provided by the Pyserini toolkit~\cite{lin2021pyserini}.

\textbf{Baselines.}
We compare \method{} against two categories of approaches: OCR-based text retrievers and visual retrievers. For text retrievers, we first extract textual content from each document image using PaddleOCR~\cite{cui2025paddleocr} and perform retrieval over the resulting OCR text. These text retrievers consist of BM25~\cite{robertson2009probabilistic}, a lexical matching method; BGE~\cite{xiao2024c}, a strong dense text retriever; E5-Mistral-7B-Instruct~\cite{wang2024improving} and NV-Embed~\cite{lee2025nv}, which are powerful LLM-based embedding models. For visual retrievers, we evaluate CLIP~\cite{radford2021learning}, a dual-encoder vision-language model, and SigLIP 2~\cite{tschannen2025siglip}, a contrastive vision-language pretraining model with a sigmoid loss, as well as VLM-based retrievers such as VLM2Vec~\cite{jiang2025vlm2vec} and E5-V~\cite{jiang2024e5}. We also consider visual document retrievers that are specifically optimized for visual document retrieval, including DSE~\cite{ma2024unifying}, ColPali~\cite{faysse2025colpali}, and VDocRetriever~\cite{tanaka2025vdocrag}. For VDocRetriever, we report results based on our reproduction using its official implementation with settings aligned to \method{}. For all other models, we use their official checkpoints.

\begin{table}[t]
\small
\centering
\caption{Prompt Templates Used to Prompt VLMs to Generate Region-Focused Descriptions.}
\label{tab:prompt}
\renewcommand{\arraystretch}{1.12}
\setlength{\tabcolsep}{8pt}
\begin{tabular}{|p{\dimexpr0.96\linewidth-2\tabcolsep-2\arrayrulewidth\relax}|}
\hline
\rowcolor{blue!3}\textbf{Prompt Template for VLMs to Generate Description} \\ \hline
\textbf{Task}: Given an image and a question, think step by step to find regions containing all evidence needed to answer. Each crop must be self-contained—able to answer the query on its own. When unsure, use larger boxes to ensure completeness and readability.

\textbf{Region-selection guidelines}:
1. Fully cover key evidence plus immediate context; do not clip text, numbers, or symbols.
2. Prefer complete information units (full words/lines; entire signs/labels; for charts include legend, axes, units, titles/notes).
3. Tables: include the header and relevant rows/columns with necessary context; avoid single-cell crops.
4. If evidence spans multiple parts, use multiple boxes—or one larger box if they’re adjacent.
5. Images/illustrations: include nearby numeric values or captions required by the question.

\textbf{Output format}: \{ "think": "your step-by-step reasoning", "boxes": [\{ "area": [x1, y1, x2, y2], "description": "a description of this region and why it is relevant" \}]\}

\textbf{Query}: \{ query \}\\ \hline

\end{tabular}
\end{table}

\textbf{Implementation Details.}
We use a locally deployed instance of Qwen2.5-VL-72B-Instruct~\cite{bai2025qwen2} on four A800 (40GB) GPUs to generate reasoning-guided visual cues, a process taking approximately 100 hours, following the prompt templates described in Table~\ref{tab:prompt}. During training, the retriever is initialized from Phi3V-4B~\cite{abdin2024phi3technicalreporthighly} and Qwen2.5-VL-7B-Instruct~\cite{bai2025qwen2}. All models are trained for five epochs using the AdamW optimizer with an effective batch size of 256. The training of \method{} follows a linear learning-rate decay schedule with a warmup ratio of 0.1 and a peak learning rate of $1\mathrm{e}{-4}$. We employ in-batch negatives during training, using 63 negatives per instance. The relative weight $\lambda$ of the reasoning-guided alignment loss is set to 0.2, balancing its contribution against the standard contrastive retrieval loss. To improve the training efficiency, we optimize VLMs using LoRA~\cite{hu2022lora} in combination with Flash Attention~\cite{dao2024flashattention}.

\begin{table*}[t]
\small
\centering
\caption{Overall Performance of \method{} and Baseline Methods.
We report NDCG@5 and NDCG@10 as evaluation metrics. Some results of ColPali are omitted, as the released checkpoints are trained on data that partially overlap with the test sets.
${\dagger}$, ${\ddagger}$, ${\S}$ denote statistically significant improvements over NV-Embed$^{\dagger}$, DSE$^{\ddagger}$, and VDocRetriever$^{\S}$, respectively.}
\label{tab:overall}

\begin{tabular*}{0.98\textwidth}{l@{\extracolsep{\fill}}*{7}{rr}}
\hline
\multirow{2}{*}{Method}
& \multicolumn{2}{c}{DocVQA}
& \multicolumn{2}{c}{InfoVQA}
& \multicolumn{2}{c}{ChartQA}
& \multicolumn{2}{c}{SlideVQA}
& \multicolumn{2}{c}{PlotQA}
& \multicolumn{2}{c}{ArXivQA}
& \multicolumn{2}{c}{Average} \\
& @5 & @10
& @5 & @10
& @5 & @10
& @5 & @10
& @5 & @10
& @5 & @10
& @5 & @10 \\
\hline
\rowcolor{black!5}\multicolumn{15}{l}{\emph{Text-based retrievers}}\\
BM25        &75.6  &76.7  &39.9  &42.8  &50.0  &52.3   &49.8  &52.1 &4.4  &5.7 &33.6  &34.9  &42.2  &44.1  \\
E5-Mistral  &71.8  &73.7  &68.5  &70.7  &73.6  &74.2  &75.7  &77.4  &5.6  &6.5  &42.1  &43.3  &56.2  &57.6  \\
BGE         &70.0  &71.9  &59.4  &61.6  &61.4  &62.7  &62.4  &64.6  &4.7  &5.2  &32.4  &33.2  &48.4  &49.9  \\
NV-Embed &76.6  &78.4  &70.3  &72.5  &79.4  &80.1  &76.4  &78.5    &6.8  &7.8  &42.6  &44.1 &58.7  &60.2  \\
\hline
\rowcolor{black!5}\multicolumn{15}{l}{\emph{Multi-modal retrievers}}\\
CLIP        &29.3  &32.5  &36.1  &38.9  &33.0  &35.4  &32.2  &35.0  &9.4  &12.1  &22.6  &23.5  &27.1  &29.6  \\
SigLIP 2    &53.7  &56.2  &41.6  &44.8  &69.9  &71.9  &40.9  &43.6  &38.0  &41.9  &37.2  &39.3  &46.9  &49.6  \\
VLM2Vec     & 40.1 &42.8  & 46.8 &50.1  & 69.0 &71.4  & 44.7 &48.1  & 36.2 &39.2  & 39.5 &42.0  &46.1  &48.9  \\
E5-V      & 62.0 &63.9  & 38.2 &40.6  & 78.6 &79.9  & 59.0 &62.0  & 39.0 &43.4  & 40.9 &42.9  &53.0  &55.5  \\
DSE         & 69.0 &70.5  & 65.9 &67.8  & 76.6 &77.1  & 66.8 &69.1  & 57.6 &60.2  & 62.7 &64.0  &66.4  &68.1  \\
ColPali     & /    & /     & 62.0 &64.1  & 83.8 &84.7  & 79.0 & 80.6 & 59.1&62.2  & /    & /     & /  & / \\
VDocRetriever
            & 75.2 & 76.9
            & 72.7 & 74.9
            & 86.0 & 87.1
            & 77.2 & 78.8
            & 59.7 & 62.9
            & 69.6 & 70.8
            & 73.4 & 75.2  \\

\method{} (Phi3V)
&80.0\rlap{$^{\dagger \ddagger \S}$} &81.7\rlap{$^{\dagger \ddagger \S}$} &76.9\rlap{$^{\dagger \ddagger \S}$}  &78.6\rlap{$^{\dagger \ddagger \S}$}  &87.9\rlap{$^{\dagger \ddagger \S}$} &88.4\rlap{$^{\dagger \ddagger \S}$} &77.5\rlap{$^{\dagger \ddagger}$}  &79.5\rlap{$^{\dagger \ddagger}$}  &59.9\rlap{$^{\dagger \ddagger}$}  &63.0\rlap{$^{\dagger \ddagger}$}  &70.3\rlap{$^{\dagger \ddagger}$}  &71.8\rlap{$^{\dagger \ddagger}$}  &75.4\rlap{$^{\dagger \ddagger\S}$}    &77.2\rlap{$^{\dagger \ddagger\S}$}  \\

\method{} (Qwen)
&\textbf{86.5}\rlap{$^{\dagger \ddagger \S}$}	&\textbf{87.4}\rlap{$^{\dagger \ddagger \S}$}
&\textbf{78.6}\rlap{$^{\dagger \ddagger \S}$}	&\textbf{80.3}\rlap{$^{\dagger \ddagger \S}$}	&\textbf{93.6}\rlap{$^{\dagger \ddagger \S}$}   &\textbf{94.0}\rlap{$^{\dagger \ddagger \S}$}	&\textbf{82.5}\rlap{$^{\dagger \ddagger \S}$}	&\textbf{83.9}\rlap{$^{\dagger \ddagger \S}$}	&\textbf{62.2}\rlap{$^{\dagger \ddagger \S}$}	&\textbf{65.1}\rlap{$^{\dagger \ddagger \S}$}	&\textbf{76.2}\rlap{$^{\dagger \ddagger \S}$}	&\textbf{77.3}\rlap{$^{\dagger \ddagger \S}$}	&\textbf{80.0}\rlap{$^{\dagger \ddagger \S}$}	&\textbf{81.3}\rlap{$^{\dagger \ddagger \S}$}\\

\hline
\end{tabular*}
\end{table*}

\section{Evaluation Results}
In this section, we first evaluate the retrieval effectiveness of \method{}. We then conduct ablation studies to examine the contribution of each component within \method{}.
Furthermore, we analyze the quality of the reasoning-guided supervision signals and provide in-depth investigations of embedding space and the attention patterns of \method{} to better understand how reasoning-guided supervision enhances retrieval performance. Finally, we present case studies to further illustrate the behavior of \method{}.

\begin{table*}[t]
\centering
\small
\caption{Ablation Study of \method{}. We report NDCG@5 and NDCG@10 scores of different models. ${\dagger}$ and ${\ddagger}$ denote statistically significant improvements over the InfoNCE$^{\dagger}$ and w/o Reasoning$^{\ddagger}$ retrievers, respectively.}

\label{tab:ablation_two_backbones}
\begin{tabular*}{\textwidth}{l@{\extracolsep{\fill}}*{7}{rr}}
\hline
\multirow{2}{*}{Method}
& \multicolumn{2}{c}{DocVQA}
& \multicolumn{2}{c}{InfoVQA}
& \multicolumn{2}{c}{ChartQA}
& \multicolumn{2}{c}{SlideVQA}
& \multicolumn{2}{c}{PlotQA}
& \multicolumn{2}{c}{ArXivQA}
& \multicolumn{2}{c}{Average} \\
& @5 & @10
& @5 & @10
& @5 & @10
& @5 & @10
& @5 & @10
& @5 & @10
& @5 & @10 \\

\hline
\rowcolor{black!5}\multicolumn{15}{l}{\emph{Phi3V}}\\

InfoNCE &67.4    &69.7    &68.7    &70.8    &83.6    &85.1   &70.8    &73.2    &54.5    &58.1    &59.3    &61.3    &67.4    &69.7  \\
\method{}
& \textbf{71.5}\rlap{$^{\dagger \ddagger}$}      & \textbf{73.3}\rlap{$^{\dagger \ddagger}$}     & \textbf{72.6}\rlap{$^{\dagger}$}   & \textbf{74.7}\rlap{$^{\dagger}$}    & 85.1\rlap{$^{\dagger}$}    &86.3\rlap{$^{\dagger}$}    &\textbf{74.4}\rlap{$^{\dagger}$}     &\textbf{76.5}\rlap{$^{\dagger}$}    &\textbf{57.7}\rlap{$^{\dagger \ddagger}$}  &\textbf{61.0}\rlap{$^{\dagger \ddagger}$} &\textbf{67.7}\rlap{$^{\dagger \ddagger}$}      &\textbf{68.8}\rlap{$^{\dagger \ddagger}$}   &\textbf{71.5}\rlap{$^{\dagger \ddagger}$}    &\textbf{73.4}\rlap{$^{\dagger \ddagger}$} \\
w/o Reasoning
&67.4      &69.5     &71.9      &73.8     &\textbf{85.6}      &\textbf{86.5}    &73.6   &75.9 &55.9      &59.0   &61.3      &62.9       &69.3      & 71.3   \\

\hline
\rowcolor{black!5}\multicolumn{15}{l}{\emph{Phi3V w/ Pre-training}}\\

InfoNCE & 75.9  & 77.3  & 74.7  & 75.6  & 87.8  & 88.6  & 75.5  & 77.2  & 58.4  & 61.5  & 69.4  & 70.9   & 73.4     & 75.2  \\
\method{}
& \textbf{80.0}\rlap{$^{\dagger \ddagger}$}  & \textbf{81.7}\rlap{$^{\dagger \ddagger}$}  & \textbf{76.9}\rlap{$^{\dagger}$}  & \textbf{78.6}\rlap{$^{\dagger}$}  & 87.9  & 88.4  & 77.5\rlap{$^{\dagger}$}  & 79.5\rlap{$^{\dagger}$}  &\textbf{59.9}\rlap{$^{\dagger \ddagger}$}  &\textbf{63.0}\rlap{$^{\dagger \ddagger}$}  & \textbf{70.3}\rlap{$^{\ddagger}$}  & \textbf{71.8}\rlap{$^{\ddagger}$}  &  \textbf{75.4}\rlap{$^{\dagger \ddagger}$}     &  \textbf{77.2}\rlap{$^{\dagger \ddagger}$}  \\
w/o Reasoning
&74.5      &76.6     &76.8      & \textbf{78.6}     & \textbf{88.8}     & \textbf{89.2}     & \textbf{78.3}     &\textbf{79.8}  &58.1      &61.5   &66.8      &68.3      & 73.9     & 75.7  \\

\hline
\rowcolor{black!5}\multicolumn{15}{l}{\emph{Qwen2.5-VL-7B-Instruct}}\\

InfoNCE &79.7	&80.8	&73.2	&75.6	&92.8	&93.0	&75.6	&77.5	&57.7	&61.0	&70.5	&71.6	&74.9	&76.6  \\

\method{}
&\textbf{86.5}\rlap{$^{\dagger \ddagger}$}	&\textbf{87.4}\rlap{$^{\dagger \ddagger}$}
&\textbf{78.6}\rlap{$^{\dagger \ddagger}$}	&\textbf{80.3}\rlap{$^{\dagger \ddagger}$}	&\textbf{93.6}\rlap{$^{\ddagger}$}  &\textbf{94.0}\rlap{$^{\ddagger}$}	&\textbf{82.5}\rlap{$^{\dagger \ddagger}$}	&\textbf{83.9}\rlap{$^{\dagger \ddagger}$}	&\textbf{62.2}\rlap{$^{\dagger \ddagger}$}	&\textbf{65.1}\rlap{$^{\dagger \ddagger}$}	&\textbf{76.2}\rlap{$^{\dagger \ddagger}$}	&\textbf{77.3}\rlap{$^{\dagger \ddagger}$}	&\textbf{80.0}\rlap{$^{\dagger \ddagger}$}	&\textbf{81.3}\rlap{$^{\dagger \ddagger}$}\\

w/o Reasoning &79.5	&81.0	&76.3	&78.1	&91.2	&92.1	&79.5	&81.3	&58.4	&61.9	&69.4	&70.5	&75.7	&77.5  \\

\hline
\end{tabular*}

\end{table*}

\subsection{Overall Performance}\label{overall_performance}
Table~\ref{tab:overall} reports the overall retrieval performance of \method{} and baseline methods across six visual document retrieval benchmarks. We report statistically significant improvements using the paired t-test ($p < 0.05$).

Overall, \method{} consistently achieves substantial improvements across all six benchmarks, delivering an average performance gain of over 2\%, which demonstrates its effectiveness. By explicitly aligning representations with reasoning-guided, query-aware descriptions, \method{} enables the retriever to more accurately localize and aggregate sparse, query-relevant evidence. 
Notably, \method{} maintains significant gains across different backbone VLMs, including Phi3V and Qwen2.5-VL-Instruct, highlighting its strong generalization capability.
These results indicate that incorporating reasoning-based evidence localization and aggregation is essential for advancing visual document retrieval, rather than relying solely on stronger visual encoders and document-specific pretraining.

As shown in the results, \method{} significantly outperforms these OCR-based retrieval models by more than 17\%, demonstrating its strong effectiveness. Notably, OCR-based retrieval models typically achieve competitive performance compared to VLM-based methods on text-centric benchmarks such as DocVQA and InfoVQA. However, their performance degrades substantially on benchmarks involving complex layouts, charts, or mixed visual-textual content. This observation highlights a fundamental limitation of OCR-based pipelines: they rely solely on transcribed text, making them vulnerable to recognition errors while discarding visual cues that are crucial for evidence-oriented retrieval in visually rich documents.
In contrast, when compared with VLM-based document page retrievers that explicitly encode layout semantics for document page representations, such as DSE and VDocRetriever, \method{} significantly outperforms these models. This result indicates that \method{} is able to provide more fine-grained supervision, thereby enabling VLMs to learn more effective visual document representations.

\subsection{Ablation Study}
\label{ablation_study}
In this subsection, we present ablation studies to assess the effectiveness of the proposed reasoning-guided alignment mechanism in \method{} and to examine the sensitivity of the model to the hyperparameter $\lambda$, which controls the trade-off between the reasoning-guided alignment loss and the standard contrastive training loss commonly used in VLM training.

\textbf{Effectiveness of Components of \method{}.} As shown in Table~\ref{tab:ablation_two_backbones}, we conduct ablation studies to further assess the effectiveness of the reasoning-guided alignment strategy adopted in \method{}. Specifically, we implement \method{} on three foundation models, including Phi3V~\cite{abdin2024phi3technicalreporthighly}, Phi3V w/ Pre-training~\cite{tanaka2025vdocrag}, and Qwen2.5-VL-7B-Instruct~\cite{bai2025qwen2}. Among them, Phi3V w/ Pre-training is additionally pretrained on query-visual document pairs. In addition, we compare two ablation variants: an InfoNCE model and \method{} w/o Reasoning. The InfoNCE retriever refers to a model trained solely with the contrastive loss, without any auxiliary supervision signals. \method{} w/o Reasoning denotes the variant in which retriever training is guided by full document image captions rather than reasoning-guided descriptions.

As shown in Table~\ref{tab:ablation_two_backbones}, the full \method{} consistently outperforms the InfoNCE-trained retriever with statistically significant improvements.
Moreover, compared with InfoNCE, \method{} maintains significant gains across different backbone VLMs, including Phi3V and Qwen2.5-VL-Instruct, highlighting its robustness and strong generalization capability across different model architectures.
In contrast, removing the reasoning-guided data synthesis component from \method{} results in consistent performance degradation, particularly on benchmarks such as DocVQA, SlideVQA, PlotQA, and ArXivQA, which require identifying sparse and distributed query-relevant evidence across multiple regions.
This observation indicates that the performance gains cannot be attributed solely to the additional visual document verbalization supervision generated by VLMs. Instead, the finer-grained image descriptions are produced through query-aware reasoning, which provides region-focused signals and encourages VLMs to become more sensitive to query-relevant regions during training.

\begin{table}[t]
\centering
\small
\caption{Sensitivity Analysis of the Reasoning-Guided Alignment Weight $\lambda$. We report the average NDCG@K and Recall@K on the test set. $\lambda=0$ indicates that the effect of the reasoning-guided alignment loss is removed.}
\label{tab:parameter}

\setlength{\tabcolsep}{10pt}
\renewcommand{\arraystretch}{1.15}

\begin{tabular}{c|cccc}
\hline
\multirow{2}{*}{$\lambda$} 
& \multicolumn{2}{c}{NDCG $\uparrow$} 
& \multicolumn{2}{c}{Recall $\uparrow$} \\
\cmidrule(lr){2-3} \cmidrule(lr){4-5}
& @5 & @10 & @5 & @10 \\
\hline
0.0 & 73.4 & 75.2 & 81.7  & 87.1  \\
0.1                & 75.1 & 76.7 & 83.4 & 88.4 \\
\rowcolor{gray!10}
0.2                & \textbf{75.4} & \textbf{77.2} & \textbf{83.5} & \textbf{88.7} \\
0.3                & 75.1 & 76.8 & 83.1 & 88.2 \\
\hline
\end{tabular}
\end{table}
\textbf{Hyperparameter Analysis.}
We further analyze the sensitivity of \method{} to the hyperparameter $\lambda$ in Eq.~\ref{eq:training_loss}, which controls the relative weight of the reasoning-guided alignment loss against the contrastive retrieval objective. 
Specifically, we conduct this experiment using \method{} (Phi3V) by varying $\lambda$ over the set $\{0.0, 0.1, 0.2, 0.3\}$, and report the average retrieval performance across all six benchmarks to evaluate the sensitivity to $\lambda$.

As shown in Table~\ref{tab:parameter}, the performance of \method{} is sensitive to the choice of the hyperparameter $\lambda$. With a small alignment weight ($\lambda=0.1$), \method{} consistently yields marginal improvements over the InfoNCE-trained retriever ($\lambda=0$). When $\lambda$ is set to a larger value $0.2$, the retrieval performance of \method{} is further improved across all benchmarks, highlighting the important role of the reasoning-guided alignment loss that uses the region-focused descriptions to better optimize VLMs to learn more effective retrieval representations. However, when $\lambda$ is further increased to $0.3$, the retrieval performance degrades noticeably on all benchmarks, likely because an excessively large alignment weight overshadows the primary contrastive objective, ultimately leading to suboptimal representation learning. Finding that $\lambda = 0.2$ strikes a balance between the primary contrastive ranking objective and the auxiliary reasoning-guided alignment loss, we adopt it as the default setting for all experiments.

\subsection{Quality Analysis of Reasoning-Guided Training Signal Synthesis via \method{}}
\label{signal_quality}
In this subsection, we analyze both the quality and diversity of the descriptions generated by \method{}. In this experiment, we treat \method{} w/o Reasoning as the baseline model. Unlike \method{}, this baseline generates visual document descriptions based on the entire visual document, without grounding them in reasoning-guided, query-aware regions.

\begin{figure}[t]
    \centering
    \begin{subfigure}[b]{0.49\linewidth}
        \centering
        \raisebox{14pt}{\includegraphics[width=\linewidth]{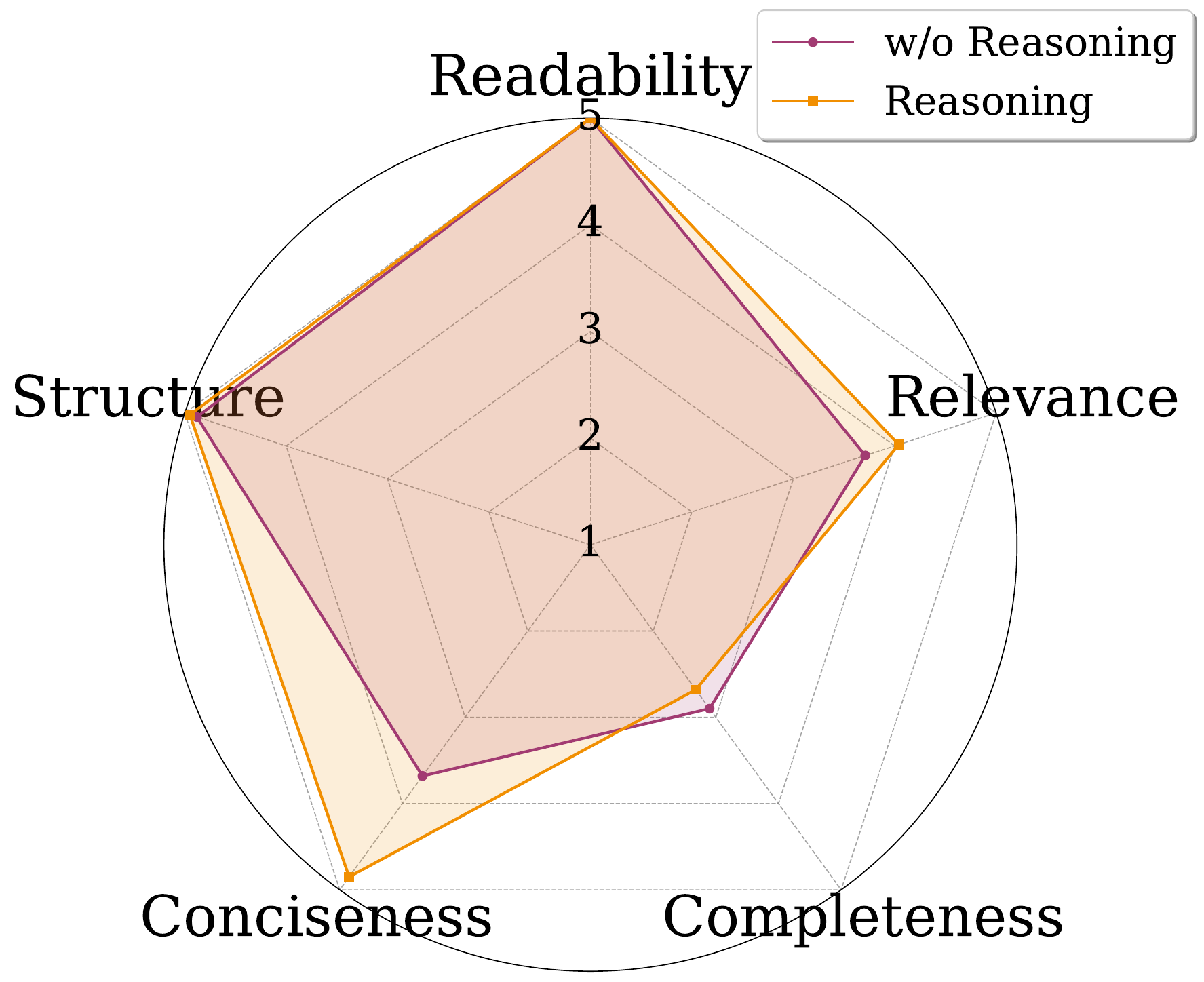}}
        \caption{LLM Evaluation Scores.}
        \label{fig:quality:llm}
    \end{subfigure}
    \hfill
    \begin{subfigure}[b]{0.47\linewidth}
         \centering
         \includegraphics[width=0.95\linewidth]{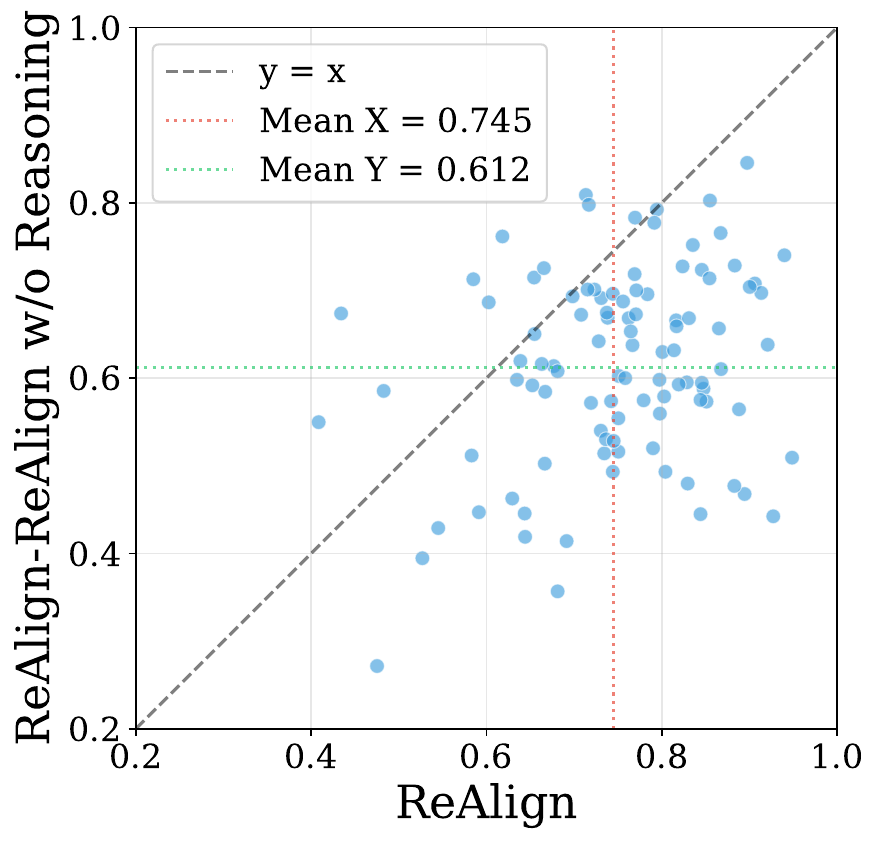}
        \caption{Similarity of Generated Descriptions.} 
        \label{fig:quality:sim}
    \end{subfigure}
    \caption{Validation of the Quality and Diversity of Supervision Signals Generated by \method{} and \method{} w/o Reasoning. Figure~\ref{fig:quality:llm} evaluates the visual document descriptions generated by \method{} using LLM-as-Judge, while Figure~\ref{fig:quality:sim} illustrates the diversity of generated descriptions.}
    \label{fig:quality}
\end{figure}

\textbf{The Quality of Region-Focused Document Description.}
To analyze the training supervision signals generated by \method{}, we randomly sample 100 examples from the training set and evaluate the quality and similarity of the visual document descriptions generated by \method{}, as shown in Figure~\ref{fig:quality}.

As shown in Figure~\ref{fig:quality:llm}, we first evaluate the quality of the visual document descriptions generated by \method{} using the LLM-as-Judge paradigm, which employs a stronger large language model, GLM-4.7~\cite{zeng2025glm}, as the evaluator.
Specifically, the GLM-4.7 model is provided with the user query and the corresponding description, and is then asked to score each query-description pair along five dimensions: readability, relevance, completeness, conciseness, and structure. The prompt template is: ``You are an expert evaluator for a RAG system. Your task is to evaluate a document image description based on a user query across five distinct dimensions\dots''.
Among the five evaluation dimensions, \method{} achieves substantially higher scores in Conciseness and Relevance, indicating that region-focused descriptions are more effective at verbalizing query-related visual cues while avoiding redundancy. In contrast, \method{} exhibits only a marginal decrease in the Completeness dimension, suggesting that focusing on query-relevant regions still preserves most of the essential information contained in the visual documents.

\begin{figure}[t]
    \centering
    \begin{subfigure}[b]{0.48\linewidth}
         \centering
         \includegraphics[width=\linewidth]{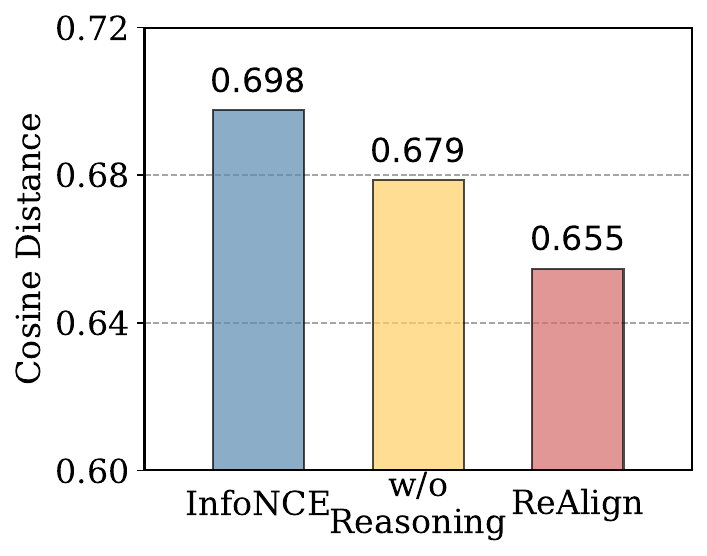}
         \caption{Query-to-Positive Distance.}
         \label{fig:embedding_space:dist_q2p}
    \end{subfigure}
    \hfill
    \begin{subfigure}[b]{0.48\linewidth}
         \centering
         \includegraphics[width=\linewidth]{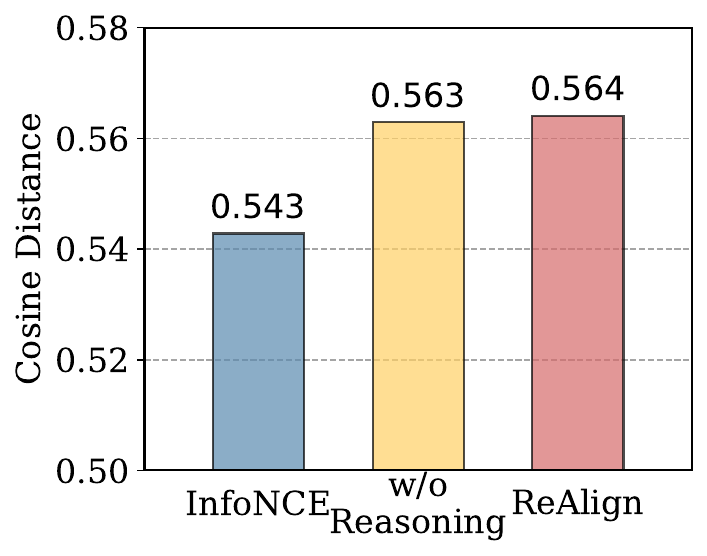}
         \caption{Document Pairwise Distance.}
         \label{fig:embedding_space:dist_pairwise}
    \end{subfigure}
    \caption{Quantitative Analysis of the Learned Embedding Space of \method{}. The quality of the embedding space is assessed by Alignment (Figure~\ref{fig:embedding_space:dist_q2p}) and Uniformity (Figure~\ref{fig:embedding_space:dist_pairwise}).}
    \label{fig:embedding_space}
\end{figure}

Furthermore, we evaluate the diversity of visual document descriptions generated by \method{}. Specifically, we utilize Qwen3-Embedding~\cite{zhang2025qwen3} to encode all generated descriptions into dense representations, and analyze both the pairwise similarity among the descriptions generated by \method{}, as well as the similarity between the descriptions generated by \method{} and those generated by \method{} w/o Reasoning. As shown in Figure~\ref{fig:quality:sim}, the results indicate that the majority of samples lie below the diagonal, suggesting that the pairwise similarity among reasoning-based descriptions is consistently higher than their similarity to the descriptions generated by \method{} w/o Reasoning. This observation demonstrates that \method{} produces descriptions with more distinct semantics from the visual documents by conditioning on the clipped regions obtained through VLM-based reasoning. In addition, descriptions generated by \method{} exhibit higher average similarity (0.745), suggesting improved semantic consistency among outputs conditioned on query-aware regions, as the VLM-based reasoning process effectively filters out irrelevant visual noise.

\textbf{The Characteristics of Learned Embedding Space.} We further conduct a quantitative analysis of the learned embedding space by randomly sampling 100 instances from the union of all testing sets. In this experiment, we assess the effectiveness of \method{} from two complementary perspectives, namely \emph{alignment} and \emph{uniformity}, as illustrated in Figure~\ref{fig:embedding_space}.

Prior studies~\cite{wang2020understanding,li2021more} have shown that contrastive learning objectives explicitly encourage both \emph{alignment} and \emph{uniformity} in the embedding space for retrieval: alignment ensures that each query is close to its corresponding positive document, while uniformity promotes a well-dispersed representation over the entire embedding space. As shown in Figure~\ref{fig:embedding_space:dist_q2p}, we first report the average cosine distance between queries and their ground-truth visual documents to assess the alignment property. The evaluation results indicate that \method{} consistently achieves lower distance values than both baseline models. This suggests that \method{} is more effective at pulling query embeddings toward their corresponding visual evidence, thereby enabling finer-grained query-document alignment. In contrast, \method{} w/o Reasoning yields query-positive distance scores closer to those of InfoNCE, indicating that descriptions generated solely from the entire visual document offer limited meaningful supervision for aligning queries with documents.
Beyond local alignment, Figure~\ref{fig:embedding_space:dist_pairwise} examines the uniformity of the embedding space by measuring the average pairwise distance among all document embeddings. The experimental results show that \method{} increases the average pairwise distance from 0.543 to 0.564. This observation suggests that \method{} not only enhances local discrimination between positive pairs but also improves the global uniformity of the representation space, thereby yielding more discriminative and well-structured document embeddings.

\begin{figure}[t]
    \centering
    \begin{subfigure}[b]{0.51\linewidth}
         \centering
         \raisebox{5pt}{\includegraphics[width=\linewidth]{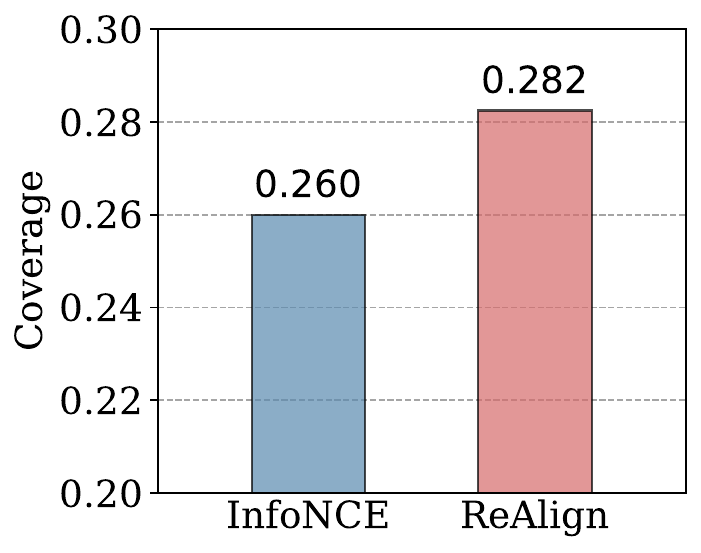}}
         \caption{Averaged Coverage Score.}
         \label{fig:attention_cover:value}
    \end{subfigure}
    \hfill
    \begin{subfigure}[b]{0.47\linewidth}
         \centering
         \includegraphics[width=\linewidth]{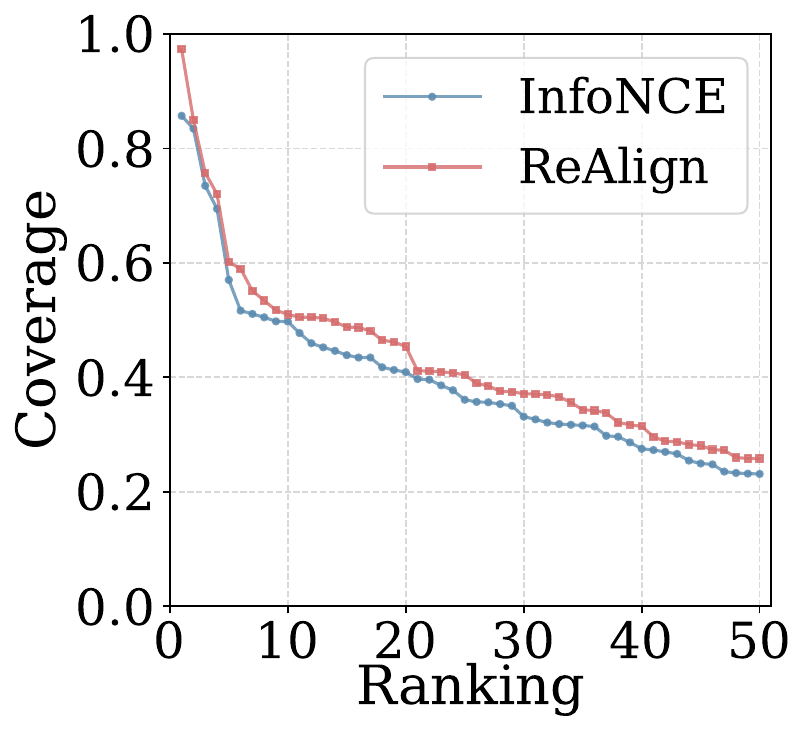}
         \caption{Coverage Score Distribution.}
         \label{fig:attention_cover:dist}
    \end{subfigure}
    \caption{Attention Coverage Score of VLMs Trained Using InfoNCE and \method{}. The coverage score is quantified as the proportion of patches within reasoning-guided clipped regions whose attention scores rank in the top 20\%.}
    \label{fig:attention_cover}
\end{figure}

\subsection{The Mechanism of \method{} of Capturing Finer-Grained Visual Cues}
In this subsection, we investigate how \method{} enables VLMs to capture finer-grained visual signals for constructing visual document representations by analyzing attention distributions of VLMs trained with InfoNCE and \method{}. In this experiment, we follow previous work~\cite{cui2025attention} to resize the visual document into crops of $336 \times 336$ pixels and then divide each crop into $28 \times 28$ patches. We treat the patch as the basic unit to show the reasoning-guided region focusing during training and the alignment between attention and document representation.

\begin{figure}[t]
    \centering
\begin{subfigure}[b]{0.51\linewidth}
     \centering
     \raisebox{12pt}{\includegraphics[width=\linewidth]{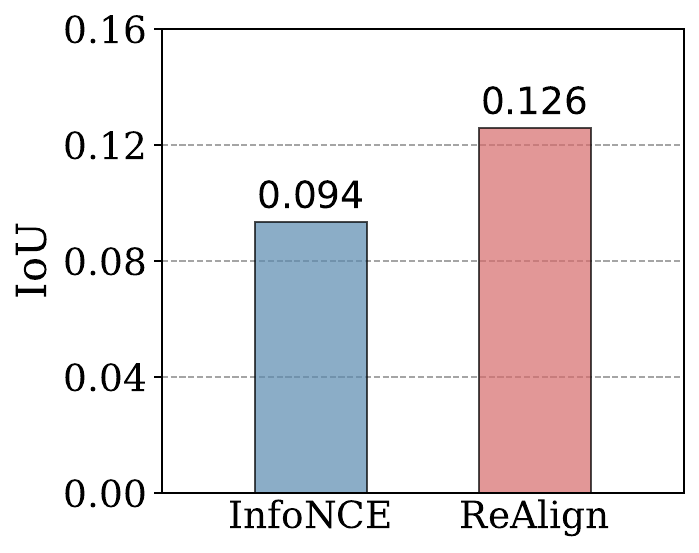}}
     \caption{Alignment IoU Score.}
     \label{fig:attn_similarity:score}
\end{subfigure}
\hfill
\begin{subfigure}[b]{0.47\linewidth}
     \centering
     \includegraphics[width=\linewidth]{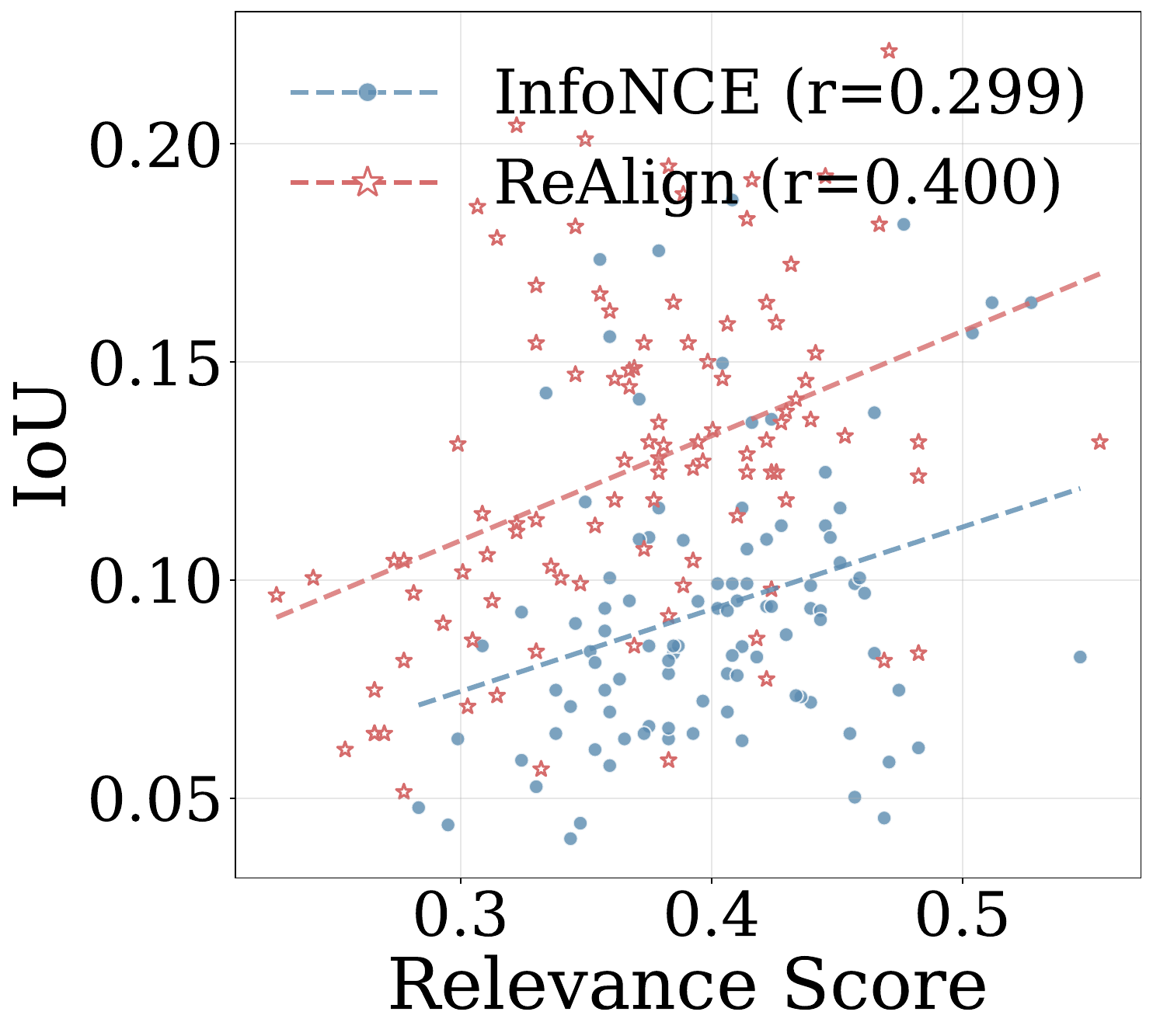}
     \caption{Correlation between Alignment IoU and Query Relevance.}
     \label{fig:attn_similarity:dot}
\end{subfigure}
\caption{Quantitative Analysis of the Alignment between VLM Attention and Visual Document Representations. The experiments evaluate the top 20\% of patches receiving the highest attention scores from the corresponding models.}
\label{fig:attn_similarity}
\end{figure}
\begin{figure*}[t]
\begin{center}
\includegraphics[width=1\linewidth]{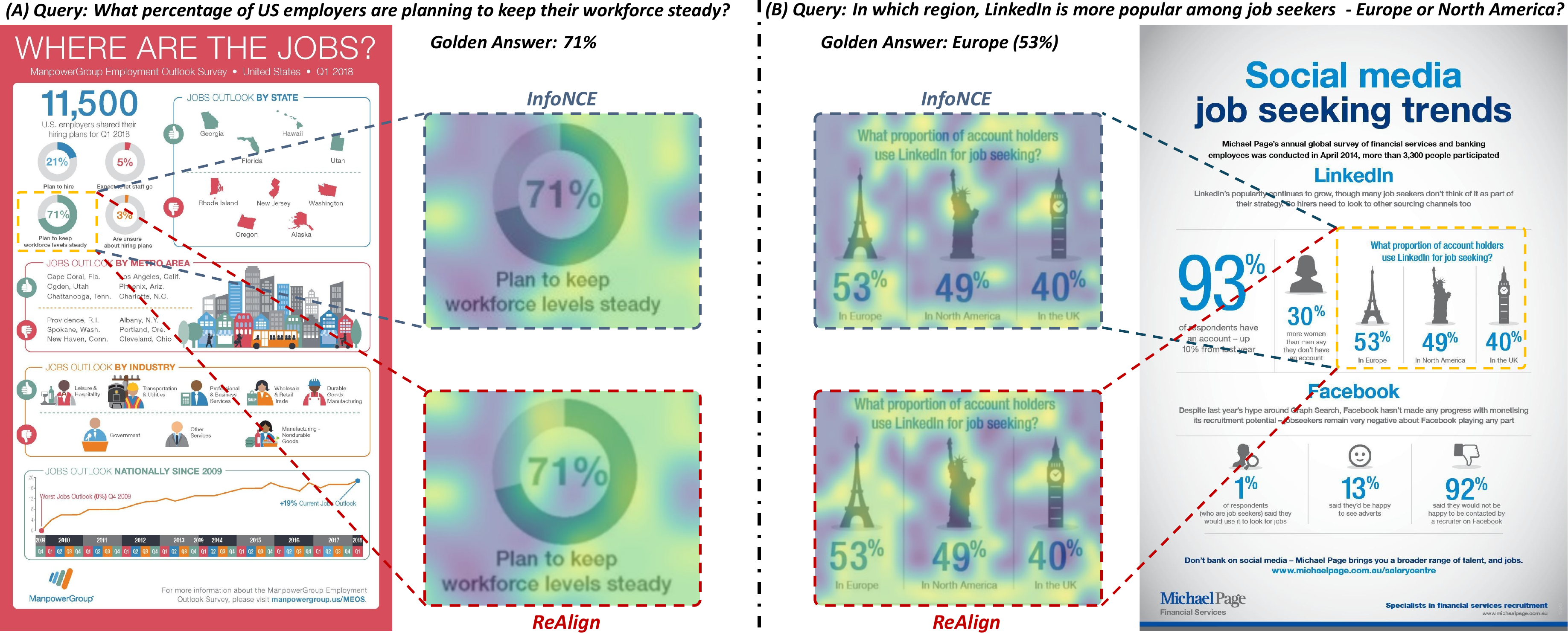}
\end{center}
\caption{Case Studies. Regions with higher color intensity indicate stronger attention.}
\label{fig:casestudy}
\end{figure*}
\textbf{Reasoning-Guided Region Focusing.}
To investigate how \method{} encourages VLMs to capture fine-grained evidence from visual documents during representation learning, we randomly sample 100 instances from the training dataset to analyze the attention variations of VLMs trained with \method{}. To quantify the alignment quality, we report the attention coverage score in Figure~\ref{fig:attention_cover}, which indicates whether the retriever is able to assign its attention to the regions selected for reasoning.

As shown in Figure~\ref{fig:attention_cover:value}, \method{} achieves a higher attention coverage over the reasoning-guided clipped regions compared to the InfoNCE training strategy, demonstrating that \method{} is able to guide VLMs to concentrate their attention on these reasoning-relevant regions, even though we only use their descriptions during training. In addition, we visualize the coverage score distribution by sorting the instances based on their attention coverage values. As illustrated in Figure~\ref{fig:attention_cover:dist}, \method{} consistently exhibits higher coverage scores than the InfoNCE baseline, indicating that \method{} can more reliably steer VLM attention toward the clipped regions of visual documents. Notably, the performance margin becomes more pronounced for the top 10\% to 50\% instances with higher coverage scores, suggesting that \method{} particularly helps the model capture more informative visual evidence in cases where VLMs trained with InfoNCE fail to confidently allocate attention over the visual page.

\textbf{Alignment between Attention and Query Relevance.}
To further investigate how \method{} enhances VLMs in retrieving fine-grained document information, we analyze the consistency between patches captured by attention weights and those identified by query-based relevance scores, as illustrated in Figure~\ref{fig:attn_similarity}. Specifically, we first randomly sample 100 instances from the test set for evaluation. We then compute the union of the top 20\% patches with the highest attention scores, representing the regions on which the model focuses, and the top 20\% patches with the highest query relevance scores, indicating the regions emphasized in the final document representations.

As shown in Figure~\ref{fig:attn_similarity:score}, we report the Intersection over Union (IoU) score to measure the overlap between the two sets of regions with high attention and high query relevance, thereby quantifying the consistency between attention and retrieval semantic learning during the encoding process. The results demonstrate that \method{} substantially improves the overlap compared to the InfoNCE baseline, indicating that the agreement between attention allocation and query-based relevance is significantly enhanced through \method{}-based training.
Furthermore, as illustrated in Figure~\ref{fig:attn_similarity:dot}, we analyze the correlation between attention and visual representations by plotting the IoU scores against query relevance scores for regions with high attention weights. The results suggest that, during training, VLMs are able to capture latent information in visual documents that is potentially relevant to downstream queries. Notably, \method{} achieves a higher correlation between IoU scores and query relevance scores than InfoNCE, demonstrating its effectiveness in strengthening the alignment between attention mechanisms and query-focused semantic signals. Benefiting from reasoning-guided, region-focused description generation, \method{} enables VLMs to more effectively capture query-relevant information during visual document representation learning.

\subsection{Case Study}
In this subsection, we conduct case studies to demonstrate the effectiveness of our \method{} model. As shown in Figure~\ref{fig:casestudy}, we sample two InfoVQA examples and visualize the attention distributions of VLMs trained with the InfoNCE objective and with \method{} over the query-relevant regions of the document pages.

For Case A, the query asks for the specific percentage of employers planning to keep their workforce steady. The relevant evidence is highly localized, where the answer is provided by a numeric value located at the center of the corresponding pie chart. However, the InfoNCE-based retriever is distracted by semantically related but non-decisive context, with its attention dispersed across surrounding descriptive text and only partially covering key regions. As a result, the model fails to capture critical visual cues from the document, such as the ``71\%'' value that directly answers the query. In contrast, the VLM optimized with \method{} allocates its attention more effectively to the golden region, covering the crucial numerical information. This observation indicates that \method{} enables VLMs to better focus on and capture essential evidence during training.
For Case B, the query requires comparing LinkedIn's popularity between Europe and North America. The VLM optimized with InfoNCE exhibits a strong attention bias toward textual content in the visual document, while exhibiting insufficient coverage of numerical information such as ``53\%'', which is directly relevant for answering the query. Such an attention pattern may cause VLMs to predominantly encode textual features while overlooking important numerical or visual cues during representation learning. 
In contrast, \method{} produces a broader and more balanced attention distribution over critical regions of the document, benefiting from its reasoning-guided region focus alignment mechanism. The attention covers both textual and numerical evidence, including ``53\%'', ``49\%'', and ``40\%''. This suggests that VLMs trained with \method{} can more effectively encode crucial information required to infer the popularity of LinkedIn across Europe, North America, and the UK, whereas VLMs trained with InfoNCE tend to focus on a single numerical value (e.g., Europe), potentially neglecting other equally important cues that play a critical role in learning robust representations.

\section{Conclusion}

In this paper, we propose \method{}, a novel framework that optimizes visual document retrieval with reasoning-guided fine-grained supervision. Our experiments demonstrate that \method{} consistently improves visual document retrievers across diverse benchmarks in both in-domain and out-of-domain settings, and generalizes well across different backbone VLMs. Further analysis shows that \method{} promotes more evidence-grounded retrieval by helping models capture fine-grained visual cues under complex document layouts.


\bibliographystyle{ACM-Reference-Format}
\bibliography{main}

\end{document}